\documentclass[aps,twocolumn,preprintnumbers,superscriptaddress,
              showpacs,nofootinbib]{revtex4}
 \newcommand{\PRE}[1]{}       
\usepackage{amsmath,amsfonts,epsfig,graphicx}
\usepackage{multirow}
\usepackage[latin1]{inputenc}
 \usepackage{amssymb}
 \usepackage{verbatim}
\usepackage{float}

\newcommand{\mgme}{{\sc MadGraph/MadEvent}}

\newcommand{\sherpa}{{\sc Sherpa}}

\newcommand{\kt}{k_T}
\newcommand{\pt}{p_T}
\newcommand{\pythia}{{\sc Pythia}}
\newcommand{\herwig}{{\sc Herwig}}

\newcommand{\Qmatch}{Q_\text{match}}

\newcommand{\ie}{{\it i.e.}}
\newcommand{\eg}{{\it e.g.}}

\begin{document}

\title{Matched~predictions~for~Higgs~production~via~heavy-quark~loops~in~the~SM~and~beyond}
 
\author{Johan Alwall}
\affiliation{Theoretical Physics Department, Fermi National Accelerator Laboratory,\\
P. O. Box 500, Batavia, IL 60510, USA}

\author{Qiang Li}
\affiliation{Paul Scherrer Institut, CH--5232 Villigen PSI, Switzerland\\
School of Physics, and State Key Laboratory of Nuclear Physics and Technology, Peking University, China}

\author{Fabio Maltoni}
\affiliation{Centre for Cosmology, Particle Physics and Phenomenology (CP3) \\
Universit\'{e} Catholique de Louvain,
Chemin du Cyclotron 2, B-1348 Louvain-la-Neuve, Belgium}

\begin{abstract}
The main Higgs production channel at hadron colliders is gluon fusion
via heavy-quark loops. We present the results of a fully exclusive
simulation of gluon fusion Higgs production based on the matrix
elements for $h + 0, 1, 2$ partons including full heavy-quark loop
dependence, matched to a parton shower. We consider a Higgs with
standard model couplings as well as models where the Higgs has
enhanced couplings to bottom quarks ($b$-philic). We study the most
relevant kinematic distributions, such as jet and Higgs $\pt$ spectra and
find that matched samples provide an accurate description of the final
state. For the SM Higgs, we confirm the excellent accuracy of the
large heavy-quark-mass approximation also in differential
distributions over all phase space, with significant effects arising
only at large $\pt$. For a $b$-philic Higgs however, the loops have a
dramatic impact on the kinematics of the Higgs as well as of the jets
and need to be accounted for exactly to achieve reliable event
simulations.
\end{abstract}


\maketitle

\section{Introduction}\label{intro}

The CERN Large Hadron Collider (LHC) is running at a center of mass energy
of 7 TeV and it has already  accumulated several inverse femtobarns of integrated luminosity per experiment.
One of its main goals is to explore the details of electroweak symmetry breaking, and
in particular to establish the existence of a Higgs sector of or beyond the Standard Model (SM).

At the LHC, Higgs boson production mainly proceeds via a quantum effect, gluon
fusion (GF)~\cite{Georgi:1977gs}. This is induced by heavy-quark loops, in 
particular the bottom and the top quarks, the latter being by far the dominant one
in the SM.  For a not too heavy Higgs boson ($m_h \lesssim 2 m_t$), and in appropriate kinematic regions
($p_T^h\lesssim m_t$), the  top quark can be integrated out, resulting,  to a
very good approximation, in a simple, non-renormalizable effective field 
theory, ${\cal L}_{\rm HEFT} = - \frac14  \frac{h}{3 \pi v} F^{\mu \nu, a} F_{\mu \nu}^{a } $
(HEFT)~\cite{Ellis:1975ap,Shifman:1979eb,Kniehl:1995tn}, $v$ being the 
Higgs field vacuum expectation value and $F^{\mu \nu,a}$  the QCD field tensor. 
The next-to-leading order (NLO) QCD corrections~\cite{Dawson:1990zj,Djouadi:1991tka,Graudenz:1992pv,Spira:1995rr}
were calculated decades ago in both HEFT as well in the full SM and 
found to be very large ($\sigma^{\rm NLO}/\sigma^{\rm LO} \sim 2$). 
This motivated the formidable endeavour of the next-to-next-to-leading order
(NNLO) QCD calculations, which have been fully evaluated in 
HEFT~\cite{Harlander:2002wh,Anastasiou:2002yz,Ravindran:2003um}. 
The exact NNLO calculation involves three loop massive diagrams, 
and is currently out of reach.  However, recently, 
the finite top-quark mass effects to the total NNLO prediction  have
been estimated through a power expansion~\cite{Harlander:2009my,Pak:2009dg,Harlander:2010wr,Pak:2011hs}
and found to have a negligible impact on total rates. Soft gluon 
resummation effects have also been studied in HEFT at NNLL~\cite{Catani:2003zt,Moch:2005ky}.   
On the other hand, it is known that in the hard tails of differential distributions or even
in special kinematics regimes, such as at small-$x$~\cite{Marzani:2008az}, 
loop effects need to be accounted for exactly. 
So far, the recommended best predictions for Higgs GF inclusive production 
rates in the standard model~\cite{Dittmaier:2011ti} are based on the 
NNLO+NNLL results in HEFT, while keeping the heavy quark mass dependence at NLO+NLL~\cite{Spira:1995mt}. 

In Beyond the SM (BSM) theories, GF becomes sensitive to all colored states in the 
spectrum with significant couplings to the Higgs(es). Even though the bookkeeping becomes more involved, 
as long as such states are heavy, an effective field approximation can still
be used and QCD corrections can be computed as in the SM.
The only genuine complication arises, not from additional heavy BSM particles, but from the possibility of  bottom quarks to have enhanced
couplings to the scalar (or pseudo-scalar) states of the theory. In
SUSY, and more generally in type II two-Higgs-doublet-model scenarios,
this corresponds to a large $\tan \beta$ scenario (where
$\beta=v_1/v_2$, $v_{1,2}$ being the vacuum expectation values of the 
Higgs doublets coupling to down- and up-type fermions, respectively).
 In this case, the HEFT approximation cannot be employed and the accuracy of the best available predictions 
goes down to NLO~\cite{Dittmaier:2011ti}.

Being of primary importance, total rates and Higgs kinematic distributions are now quite well predicted and also available via public
codes such as ResBos~\cite{Balazs:2000sz} and
HqT~\cite{Bozzi:2005wk,deFlorian:2011xf}. Differential $p_T^h$
distributions accurate to NLO can be also obtained via HIGLU~\cite {Langenegger:2006wu} as well as via HPro~\cite{Anastasiou:2009kn}, which both keep the exact bottom- and top-quarks mass loop dependence and therefore can be used also for predictions of scalar Higgs in BSM.
However, in experimental analyses, it is also crucial to get as precise predictions as possible for exclusive 
observables that involve extra jets, such as the jet $\pt$ spectra and the jet rates, 
at both parton and hadron level.  To optimize the search strategies  and in particular to curb the very large backgrounds,
current analyses both at Tevatron and at the LHC select  0-,1- and 2-jet events 
and perform independent analyses on each sample~\cite{Mellado:2007fb}. The final systematic 
uncertainties are effected by both the theoretical and experimental ones of such a jet-bin based separation, see, \eg, Ref.~\cite{Stewart:2011cf}.
In the HEFT, fully exclusive parton- and hadron-level calculations can be performed by Parton
Shower (PS)  programs such as
\pythia~\cite{Sjostrand:2006za},  \herwig~\cite{Corcella:2000bw} and
\sherpa~\cite{Gleisberg:2008ta} in the soft and collinear approximation, or with NLO QCD codes matched
with parton showers: via the MC@NLO~\cite{Frixione:2010wd} and POWHEG~\cite{powheg0,powheg1,powheg2,Alioli:2008tz} methods.
However, beyond the HEFT, no fully exclusive prediction has been available so far. 
The reason is that one needs to compromise between the validity of HEFT and the complexity of higher
loop calculations.  It is however possible to get full exclusive control
at hadron level on the complex event topology at the LHC, while still
reaching approximately NLL accuracy, with the help of recent
sophisticated matching methods between matrix elements and parton
showers~\cite{Catani:2001cc,Alwall:2007fs}.

In PS programs, QCD radiation is generated in the collinear and soft approximation, using Markov chain
techniques based on Sudakov form factors. Hard and widely separated
jets are thus poorly described in this approach. On the other hand,
tree-level fixed order amplitudes can provide reliable predictions in
the hard region, while failing in the collinear and soft limits. To
combine both descriptions and avoid double counting or gaps between
samples with different multiplicity, an appropriate matching method is
required. Several algorithms have been proposed over the years: the
CKKW method, based on a shower veto and therefore on event
re-weighting~\cite{Catani:2001cc} and MLM
schemes, based on event rejection~\cite{Mangano:2006rw,
  Alwall:2007fs}. 

In this work, we report on  the first matched simulation of Higgs
production in gluon fusion that retains the full kinematic
dependence on the heavy-quark loops, in the SM as well as in
generic scenarios with enhanced Higgs couplings with bottom quarks,
which we dub ``$b$-philic Higgs''. 

The paper is organized as follows.
We first describe our methodology. Then we present our results for
a SM Higgs. We show that the matching procedure provides reliable
results both at the Tevatron and especially at the LHC and that the effects from massive quark loops
are indeed mild over all phenomenologically relevant phase space. 
The  $b$-philic Higgs is considered in the following section, where it is
shown that loop effects must be included exactly. Gluon fusion production is also compared 
to a matrix-element matched sample for $b\bar b \to h$, which is the dominant production mode 
in this scenario. We draw our conclusions in the last section.

\section{Method}

Our study is based on the $\kt$-MLM and shower-$\kt$ matching
schemes implemented in \mgme~\cite{Alwall:2007st}, interfaced with
\pythia~6.4 for parton shower and hadronization. As explained below, 
we find it convenient to include the effects of the heavy-quark loop by simply 
reweighting the events generated via tree-level HEFT amplitudes.

In the $\kt$-jet MLM matching schemes~\cite{Alwall:2007fs,Alwall:2008qv}, matrix
element multi-parton events are produced with a minimum separation
$\kt$ cutoff of $Q^{\rm ME}_{\rm min}$. For every event, the
final-state partons are clustered according to the $\kt$ algorithm,
and the $\kt$ value for each clustering vertex corresponding to a QCD
emission is used as renormalization scale for $\alpha_S$ in that
vertex. For the central hard $2\to 1$ or $2\to 2$ process, the
transverse mass $m^2_T = p^2_T + m^2$ of the particle(s)
produced in the central process is used as factorization and
renormalization scale. Subsequently, this event is passed to the
\pythia~parton-shower generator. There, one of two schemes is
employed. Either, the final partons (after parton
showering) are clustered into jets, using the $\kt$ algorithm with a
jet cutoff of $Q^{\rm jet}_{\rm min} > Q^{\rm ME}_{\rm min}$. The jets
are considered to be matched to the original partons if $\kt {\rm
  (parton, jet)}$ is smaller than the cutoff $Q^{\rm jet}_{\rm
  min}$. If any parton is not matched to a jet, the event is
discarded. For events with parton multiplicity smaller than the
highest multiplicity, the number of jets must be equal to the number
of partons. We call this scheme the the $\kt$-MLM scheme. Alternatively,
no matching between shower jets and partons is
done. Instead, an event is retained provided that the hardest emission
in the \pythia\ parton shower is below the scale $Q^{\rm jet}_{\rm
  min}$ (or, for events from the highest multiplity, below the scale
$Q^{\rm parton}_{\rm min}$ of the softest parton in the event).  This
is called the ``shower-$\kt$'' scheme, and allows for the matching scale $Q^{\rm jet}_{\rm
  min}$
to be set equal to the matrix element cutoff scale $ Q^{\rm ME}_{\rm
  min}$. The two matching
schemes have been shown to give equivalent
results \cite{Alwall:2008qv}, but for the case of $b$-philic Higgs, the shower-$\kt$ scheme 
allows for lower matching scales and it is therefore more efficient.

In order to take into account the full kinematic dependence of the
heavy quark loop in Higgs production, the full one-loop amplitudes for
all possible subprocesses contributing to $h+0,\,1,\,2$ partons have been calculated. 
Analytic expressions have been generated with FeynArts~3.5 \cite{Hahn:2000kx}, and manipulated with
FormCalc~5.3 \cite{Hahn:1998yk}.  The tensor integrals have been  evaluated
with the help of the LoopTools-2.5 package \cite{Hahn:1998yk}, which
employs the reduction method introduced in Ref.~\cite{Denner:2002ii} for
pentagons, and Passarino-Veltman reduction for the lower point
tensors. The resulting regular scalar integrals are evaluated with the FF
package~\cite{vanOldenborgh:1990yc}. We have also implemented 
 the reduction method for pentagon tensor integrals as proposed in
Ref.~\cite{Denner:2005nn} for better numerical stability. The codes have
been used and validated against known results in a previous study~\cite{Li:2010fu}.
The final implementation of the calculation includes the contributions from bottom and top quarks
and their (destructive) interference.
The evaluation of multi-parton loop amplitudes is, in general, computationally quite expensive. 
Moreover, in the case of inclusive matched samples, an efficient event generation needs as a first
phase a rather thorough exploration of the phase space. It therefore becomes advantageous to devise
a method where the mapping of the integrand can be done in a quick (though approximate) way and the
evaluation of loops limited to a small number of points. Our strategy is as follows. 
Parton level events for  $h+0 ,\,1,\,2$ partons are generated via \mgme\ in the HEFT model, 
with scale choices optimized  for the subsequent matching procedure. Before passing them to
the PS program, events are reweigthed by the ratio of full
one-loop amplitudes over the HEFT ones, $r= |{\cal M}_{\rm LOOP}|^2/|{\cal M}_{\rm HEFT}|^2$.
The  reweighted parton-level events are unweighted, 
passed through \pythia\ and matched using the $k_T$-MLM or the shower-$k_T$
scheme. All steps are automatic. To validate the matching procedure,  the effect of changing the matching cutoff and other parameters
 such as $Q^{\rm jet}_{\rm min}$ and $Q^{\rm  ME}_{\rm min}$ on several distributions, including the $n \to n-1$ 
 differential jet rates have been extensively assessed.

Finally, we recall that even though matrix elements for up to two final states partons are included in the simulation, 
the accuracy of the overall normalization of the inclusive sample is only leading order, exactly as in a purely parton-shower result.
It is therefore legitimate and consistent to adjust the overall normalization to the best available fully inclusive prediction for the
corresponding process. To this aim, NNLO cross sections (in fact,  just NLO for a $b$-philic Higgs) at the Tevatron and the LHC 
for the scenarios described below have been obtained via publicly available codes and collected in Table~\ref{tab:xsec}. 
\renewcommand{\arraystretch}{1.7}
\begin{table}[t]
\begin{center}
\begin{tabular}{|c|ccc|}
\hline
 & \multicolumn{3}{c|}{Cross section}\\
 & Higgs mass [GeV] & Tevatron & LHC @ 7 TeV \\
\hline  
 $gg \to h$   (SM)  & 140 & 0.672 pb &12.2 pb   \\
 $gg \to h$   (SM)  & 500 & 0.003 pb &0.869 pb  \\
 $gg \to h$  ($b$-only) & 140  & 3.0 fb & 56 fb  \\
 $b\bar b \to h$   & 140 & 4.55  fb & 135 fb  \\
\hline
\end{tabular}
\end{center}
\caption{Reference values for total cross sections for Higgs  production in the SM and considering only $b$-loops,
 used for the normalization of the inclusive samples. Results have been obtained via the {\sc HNNLO}~\cite{Florian09} and {\sc bbh@NNLO}~\cite{Harlander03} codes, with $m_t=173$\,GeV, $m_b=4.6$\,GeV, $\mu_R=\mu_F=m_h$ and employing the MSTW2008NNLO pdf set~\cite{Martin:2009iq}.}
\label{tab:xsec}
\end{table}

\section{ SM Higgs production}

\begin{figure}
\includegraphics[width=8.5cm]{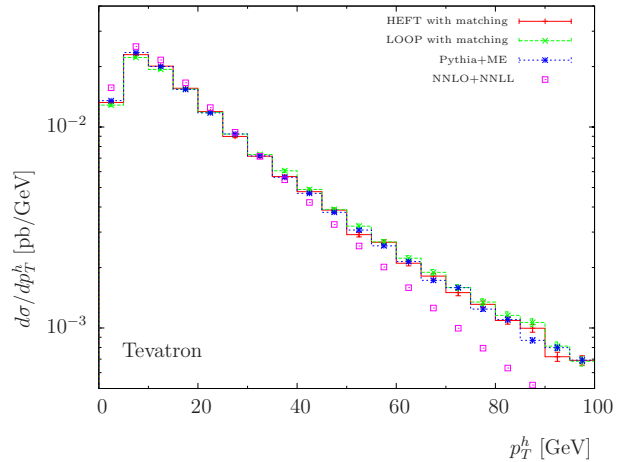}
\caption{SM Higgs  $\pt$ distributions for
$m_h=140$\,GeV in gluon fusion production at the Tevatron.  Results in the HEFT and with full loop dependence
(LOOP) are compared to the default \pythia\ implementation (which accounts for $2\to2$ matrix element corrections) and to the NNLO+NNLL results as obtained by HqT~\cite{Bozzi:2005wk,deFlorian:2011xf}. Curves normalized to the corresponding total cross sections of Table~\ref{tab:xsec}. }
\label{hpt1}
\end{figure}

\begin{figure}
\includegraphics[width=8.5cm]{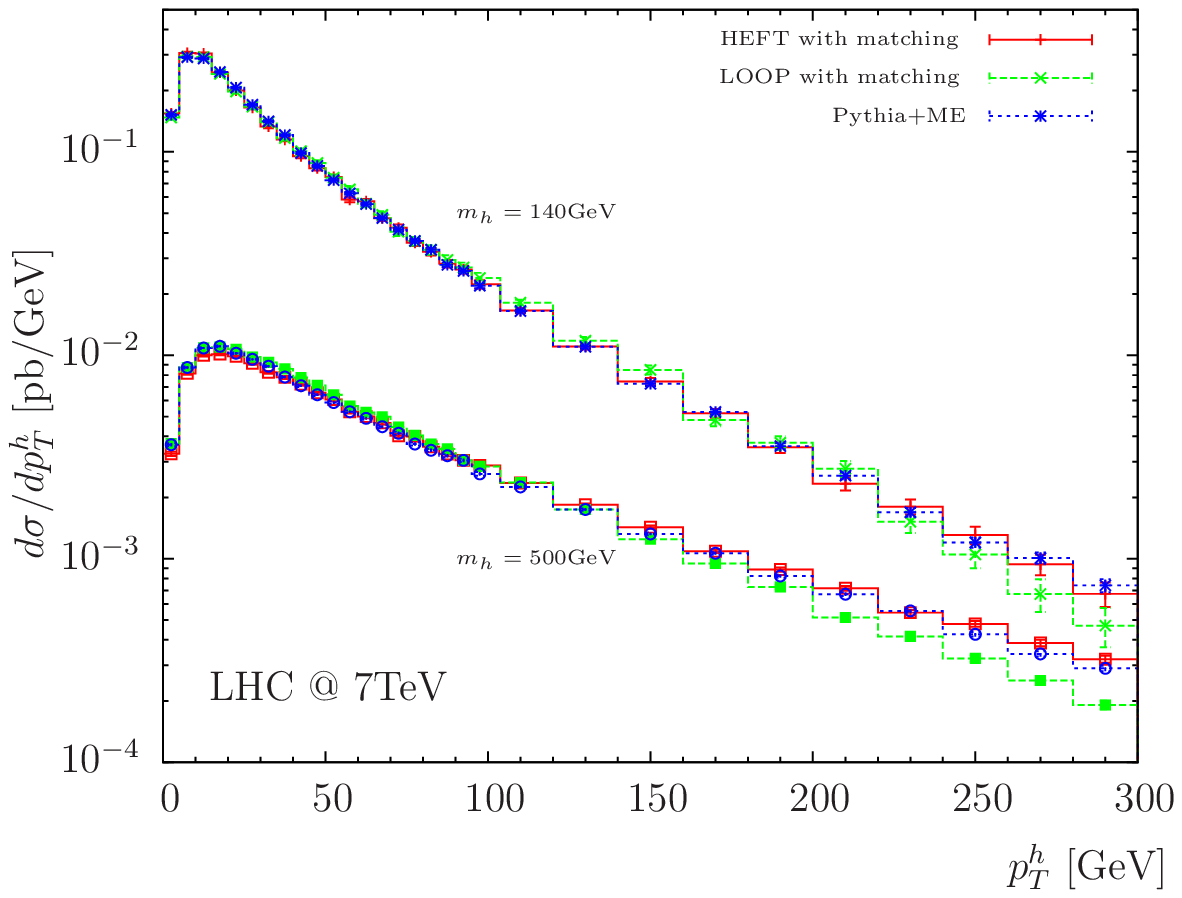}
\includegraphics[width=8.5cm]{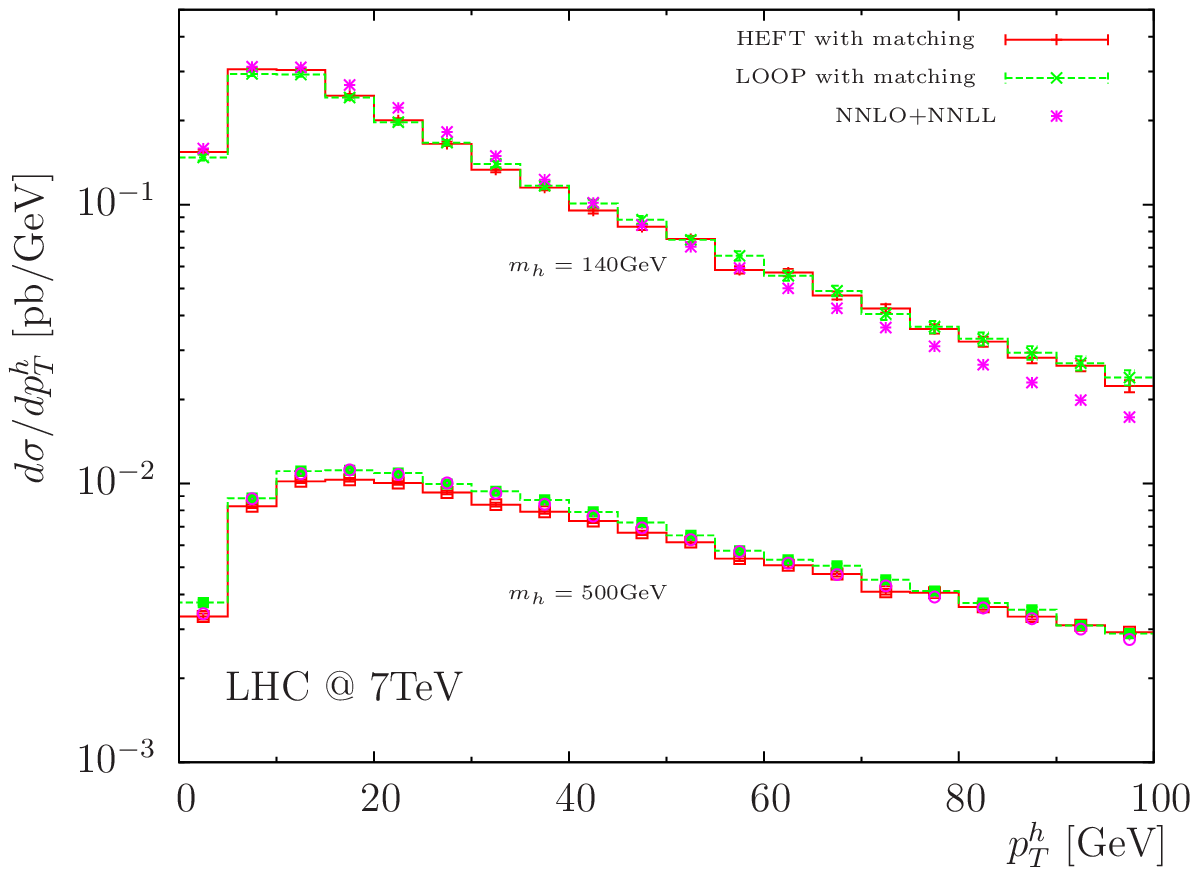}
\caption{SM Higgs  $\pt$ distributions for
$m_h=140$\,GeV and  $m_h=500$\,GeV in gluon fusion production at 7 TeV LHC.  In
the upper plot results in the HEFT and with full loop dependence
(LOOP) are compared over a large range of $p_T$ values to the default
\pythia\ implementation, which accounts for $2\to2$ matrix element corrections. In the lower plot the low-$p_T$ range is compared to the NNLO+NNLL results as obtained by HqT~\cite{Bozzi:2005wk,deFlorian:2011xf}. Curves normalized to the corresponding total cross sections of Table~\ref{tab:xsec}. }
\label{hpt}
\end{figure}

\begin{figure}
\includegraphics[width=8.5cm]{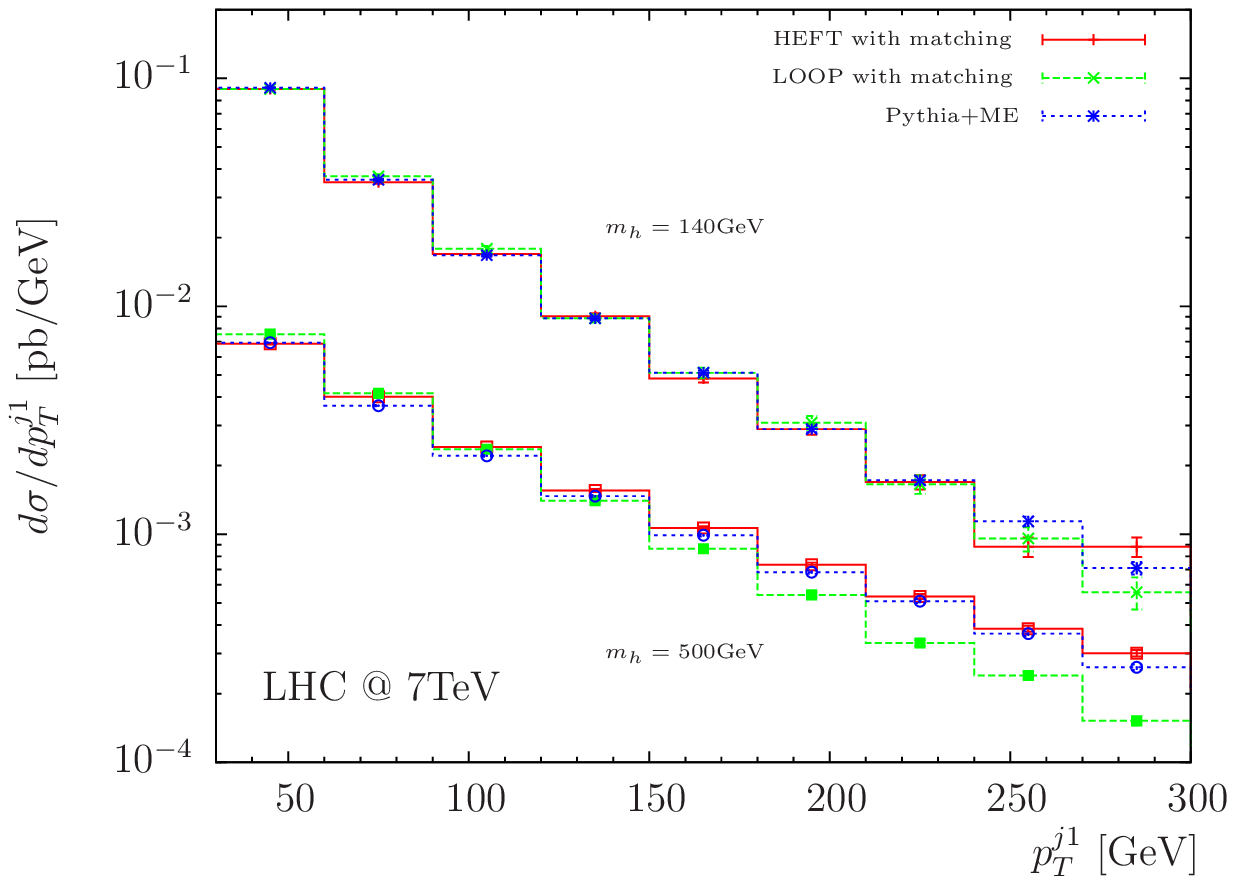}
\includegraphics[width=8.5cm]{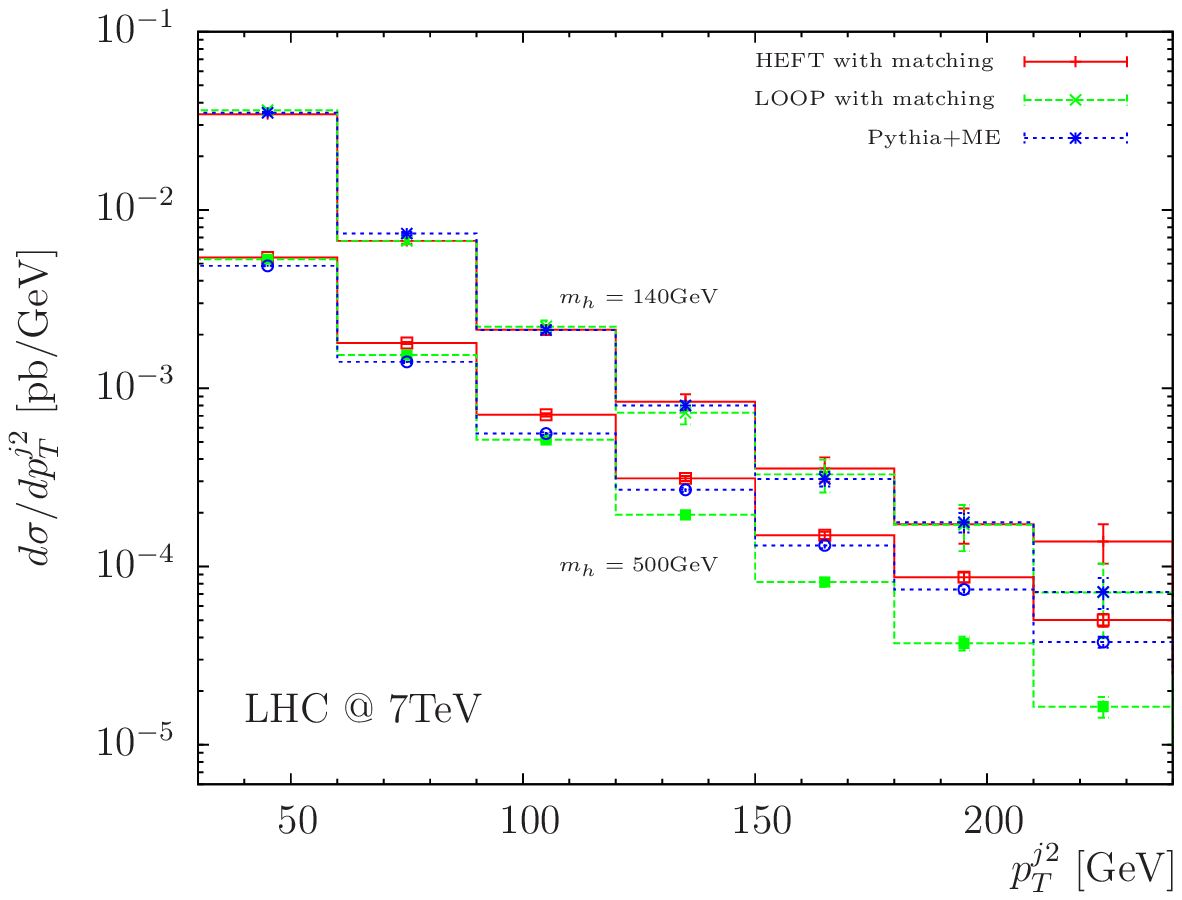}
\caption{Jet $\pt$ distributions for associated jets in gluon fusion
  production of 
$m_h=140$\,GeV and  $m_h=500$\,GeV Higgs bosons at 7 TeV LHC.} 
\label{jpt}
\end{figure}

To illustrate the results of our simulations for the Tevatron and the LHC at 7 TeV  for a standard model Higgs  
we show a few relevant observables in Figs.~\ref{hpt1}-\ref{jpt}. We define jets via the $\kt$ algorithm, with the distance measure between
parton $i$ and beam $B$, or partons $i$ and $j$ as $k_T^{i,B}\equiv
p_T^i$, $k_T^{i,j}\equiv \min{\left(p_T^i, p_T^j\right)}
\sqrt{2(\cosh\Delta y_{ij}-\cos\Delta\phi_{ij})}/D$.  
Here $y$ is the rapidity and $\phi$ is the azimuthal angle around the beam
direction. The resolution parameter is set to $D = 1$. Jets are required to satisfy $|\eta_{j}|< 4.5$ and $p_T^{j}>30\,{\rm GeV}$.
For sake of simplicity, we adopt Yukawa couplings corresponding to the
pole masses, \ie,  for the top quark  $m_t=173$\,GeV and for the bottom-quark mass $m_b=4.6$\,GeV. Other quark masses are neglected. 
Throughout our calculation, we adopt the CTEQ6L1 parton distribution functions
(PDFs)~\cite{Pumplin:2002vw} with the core process renormalization and
factorization scales $\mu_r=\mu_f=m_T^h\equiv\sqrt{(p_T^h)^2+m_h^2}$.
For the matching performed in \mgme, the 
$\kt$-MLM scheme is chosen, with $Q^{\rm ME}_{\rm min}=30$\,GeV and $Q^{\rm jet}_{\rm min}=50$\,GeV.

In Figs.~\ref{hpt1} we show the Higgs $\pt$ distribution for Standard Model Higgs GF production at
the Tevatron with $m_h=140$ in a range of $p_T$ relevant for experimental analysis.  
We compare matched results in the HEFT theory and in the full theory
(LOOP) with \pythia\ with $2\to 2$ matrix element corrections.  We also include the predictions from the analytic computation
at NNLO+NNLL as obtained by {\sc HqT}~\cite{Bozzi:2005wk,deFlorian:2011xf}. The curves are all normalized
to the NNLO+NNLL predictions.  The three Monte-Carlo based predictions agree
very well 
in all the shown range of $p_T$, suggesting that for this observable,
higher multiplicity matrix element corrections (starting from $2\to3$) and
loop effects are not important. This is the case also for jet $p_T$ distributions (not shown) in
the same kinematical range. The NNLO+NNLL prediction, on the other hand, suggests a slightly softer Higgs spectrum.
 
\begin{figure}
\includegraphics[width=8.5cm]{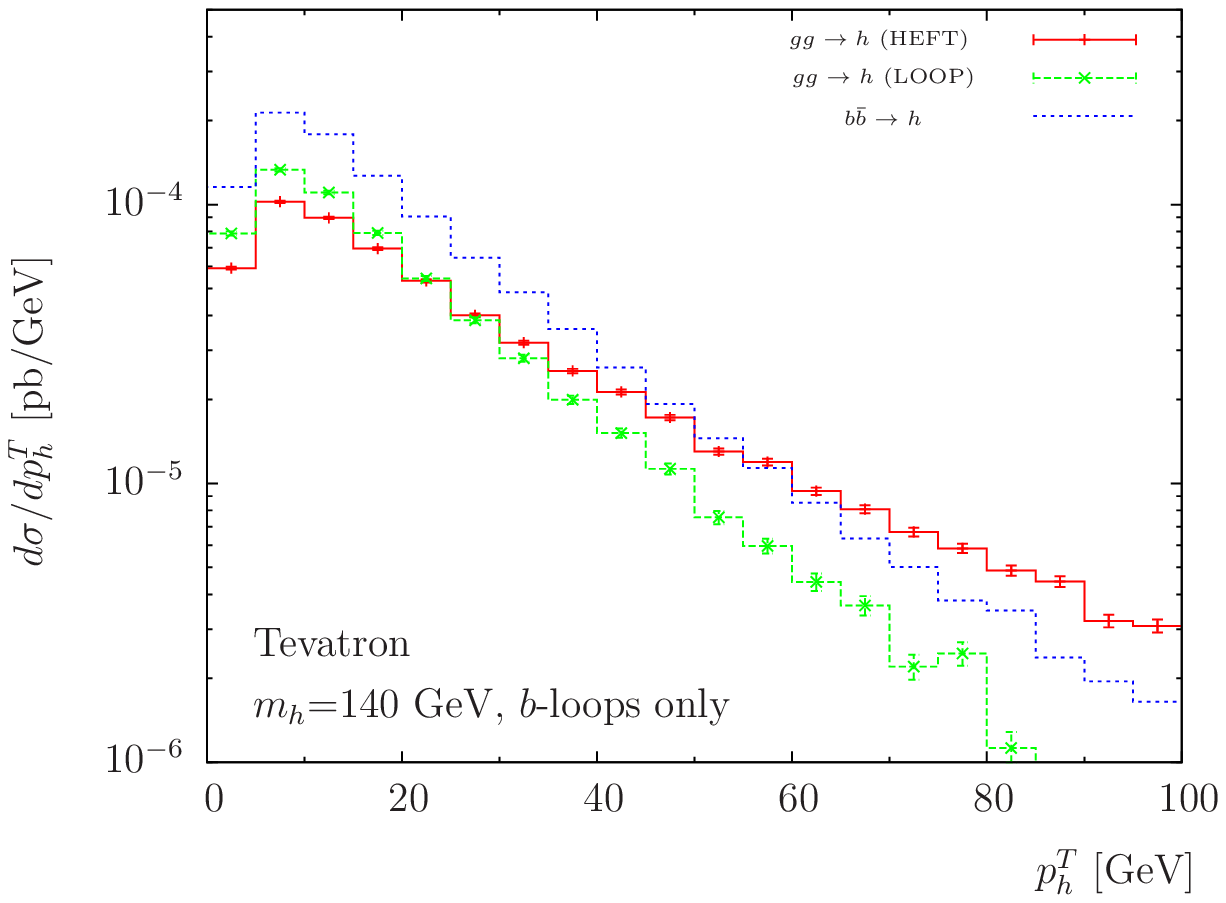}
\includegraphics[width=8.5cm]{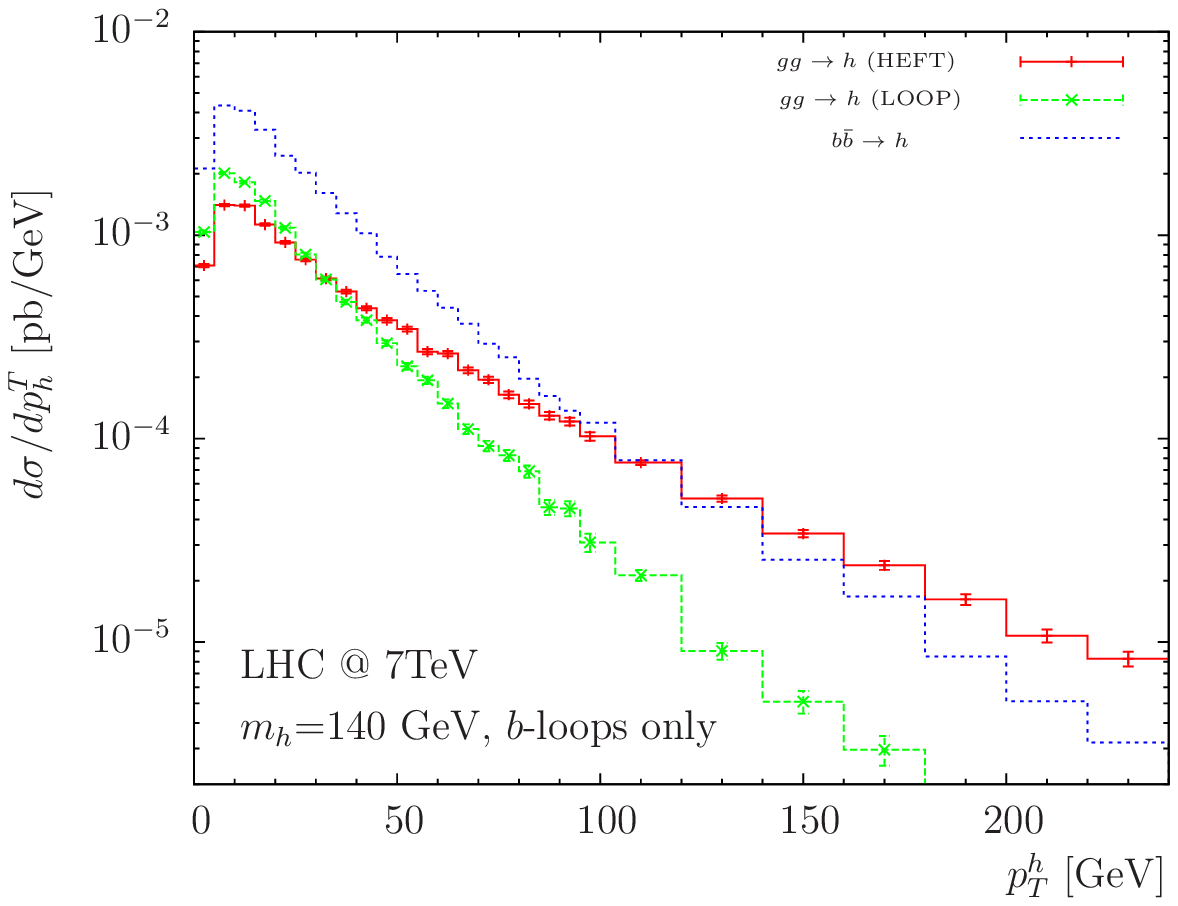}
\caption{$b$-philic Higgs  $\pt$ distribution a the Tevatron and the LHC with $m_h=140$\,GeV.   Results in the HEFT approximation (red curve) 
and with full loop dependence (green) are shown.  Spectrum of  Higgs produced via $b\bar b$ fusion in the five flavor scheme is also shown.  All samples are matrix-element matched with up to two partons in the final state.  Curves normalized to the corresponding total cross sections of Table~\ref{tab:xsec}. }
\label{mt0pth}
\end{figure}

In Figs.~\ref{hpt}-\ref{jpt}, we show the Higgs and jet $\pt$
distributions for Standard Model Higgs GF production at
the 7TeV LHC with $m_h=140$ and $500$\,GeV. 
Once again the Monte-Carlo based results agree well with each other. As expected, loop effects
show a softening of the Higgs $p_T$, but only at quite high $p_T$. 
We also see that the heavier the Higgs, the more important are the loop
effects. 
This is expected, since the heavy Higgs boson can probe the internal structure of 
the top-quark loop already at small $\pt$. The jet $p_T$ distributions do confirm the overall
picture and again indicate loop effects to become relevant only for rather high values
of the $p_T$.

The agreement, on the other hand, of the NNLO+NNLL predictions at small $p_T$ for
both Higgs masses it is quite remarkable. In this respect, our analysis strongly motivates  
the use of matched samples for simulating GF Higgs production at the LHC. Key distributions, such
as the $p_T$ of the Higgs, do agree remarkably well with the best available
predictions, for example NNLO+NNLL at small Higgs $p_T$, and offer
improved and easy-to-use predictions for other key observables such as the jet 
rates and distributions. In addition, for heavy Higgs masses and/or large $p_T$, 
loop effects, even though marginal  for phenomenology,  can also be taken 
into account in the same approach, if needed.

\section{b-philic Higgs production}

\begin{figure}
\includegraphics[width=8.5cm]{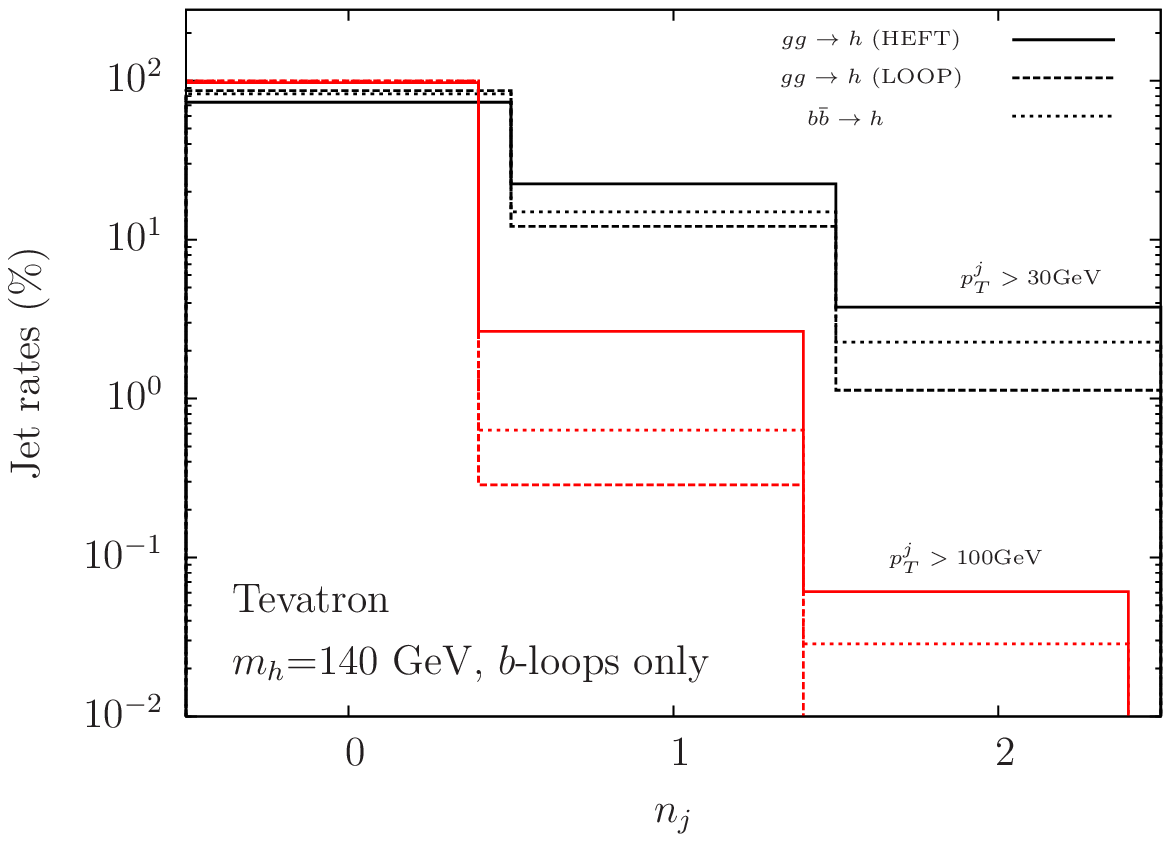}
\includegraphics[width=8.5cm]{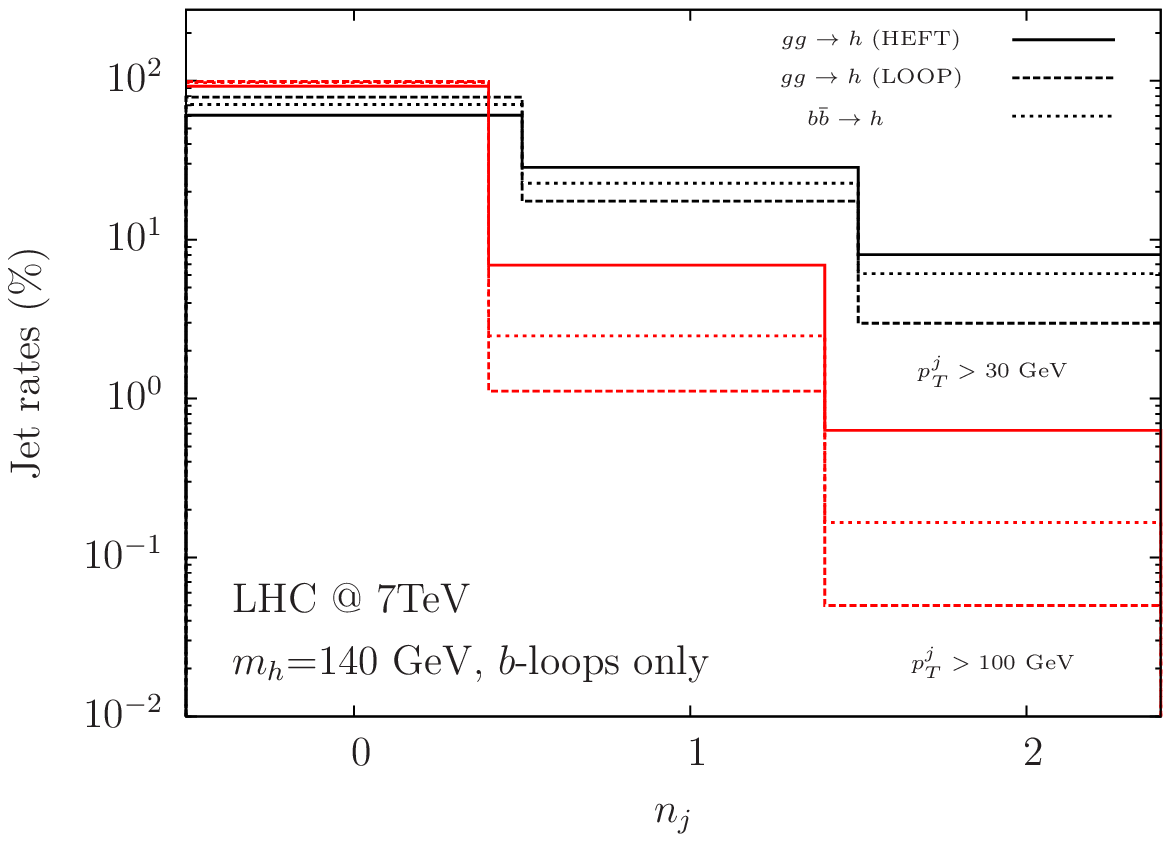}
\caption{Jet rates for $b$-philic Higgs production 
(where only the coupling to the bottom quark is included). 
$m_h=140$\,GeV at the Tevatron and LHC ($\sqrt{s}=7$ TeV). } 
\label{mt0jetrates}
\end{figure}
In this section we present the results of a simulation of a $b$-philic Higgs.
Parameters are the same as in the previous section, except that, as explained below, the
top Yukawa coupling is set to zero and  the matrix-element matching  in \mgme\  is
performed through the shower-$\kt$ matching scheme with $\Qmatch=10$\,GeV.

In Fig.~\ref{mt0pth}, we show the $p_T^h$ distributions for GF
production at the 7 TeV LHC of a $b$-philic Higgs with $m_h=140$\,GeV.
We remind the reader that in our calculation the bottom-quark and top-quark masses
can be chosen independently as well as the value of the corresponding Yukawa couplings. We
can therefore study the production of scalars with arbitrary couplings to the heavy quarks
such as those appearing in a generic two Higgs doublet model or in the minimal supersymmetric standard model.
For the sake of illustration, we define a  simplified scenario where the Higgs coupling to the top quark is set to zero.
In so doing, we study the Higgs and jet distributions relative to 
a ``large $\tan\beta$" scenario with  bottom-quark loops dominating. Note that for simplicity we keep the same normalization 
as in the standard model, i.e., $y_b/\sqrt{2} = m_b/v$ with $m_b=4.6$ GeV, as the corresponding cross sections
in enhanced scenarios can be easily obtained by rescaling. 

In the $b$-philic Higgs production, the particle running in the loop is nearly massless, and there is no
region in $m_h$ or $p_T$ where an effective description is valid. This
also means that a parton-shower generator 
alone has no possibility of correctly describing the effects of jet radiation, and genuine loop matrix-elements
plus a matched description are needed for achieving reliable simulations. 

In fact, the largest production cross section for a $b$-philic Higgs does  not come  from loop induced gluon fusion, but from tree-level $b \bar b$ fusion.
Phenomenologically, it is therefore very important to be able to also generate events for this kind of process, which typically leads to 
final states with more $b$-jets than the GF production. We do so by  matching tree-level matrix elements 
for  $h+0,1,2$ partons  (with a $hb\bar b$ vertex)  in the five flavor scheme to the parton shower.  In so doing we provide a complete  and consistent event 
simulation of inclusive Higgs production in a $b$-philic (or large
$\tan \beta$) scenario.  We note in passing that a four-flavor scheme,
\ie,  starting from the leading order process $gg \to b\bar b h$, could also be employed. While this latter approach  has some important advantages, it also offers complications with respect to the simpler five flavor scheme. A detailed comparison between the two approaches, which are known to be  compatible at the level of total cross sections (see, \eg, \cite{Dittmaier:2011ti} and references therein), would certainly be welcome.  Being beyond the scope of this paper, however, we leave it to future work.
 
Fig.~\ref{mt0jetrates} shows jet rates for the Tevatron and 7 TeV LHC for a 
$b$-philic Higgs for two minimum jet $p_T^j$ scales,
30 and 100 GeV. As is readily seen from the figure, the
effect of properly including loop effects is significant already with
a jet $p_T^j$ cutoff at 30 GeV, and increasingly important for larger
cutoff values. This immediately translates to the effect of a jet veto
with a given $p_T^j$ cutoff for the veto.

\section{Conclusions}

In summary, we have presented the first fully exclusive simulation
of gluon fusion inclusive Higgs production based on the exact one-loop
matrix elements for $h+0,\,1,\,2$ partons, matched to \pythia\
parton showers using multiple matching schemes implemented in \mgme.
We have compared the loop reweighted matched results 
with the corresponding  HEFT results, \pythia\ results, and, when possible,  with NNLO+NNLL predictions. 
We have considered both the SM Higgs and the case of scalar particles with 
enhanced couplings to bottom quarks and studied the most relevant 
kinematic distributions, such as jet and Higgs $\pt$ spectra.
Our results highlight the relevance of a complete loop calculation  at large $\pt$ for a
standard model Higgs and in all phase space for $b$-philic Higgs.
Such improved simulations might be particularly relevant in searches 
performed via multivariate analysis  techniques where
details about the kinematic distributions of the Higgs decay products
and accompanying jets can have significant impact on the results.

We conclude by stressing that the method employed in this work, {\it i.e.}, 
using tree-level amplitudes based on an effective theory to generate parton-level events
and then reweighting them by the exact loop amplitudes before matching to the shower, is 
completely general and can therefore be applied to any loop-induced process. Work   
towards the automatization of this approach in {\sc MadGraph~5}~\cite{Alwall:2011uj}
via {\sc MadLoop}~\cite{Hirschi:2011pa} is in progress.

\acknowledgments
We thank Paolo Torrielli, Simon de Visscher, Massimiliano Grazzini and Giuseppe Degrassi for helpful discussions.
This work is supported by the European Community's Marie-Curie Research Training Network HEPTOOLS under
contract MRTN-CT-2006-035505, US DOE Contract No.~DE-AC02-07CH11359, by the National Natural Science Foundation
of China, by the IAP Program, BELSPO P6/11-P  and the IISN convention 4.4511.10.

\bibliographystyle{ieeetr}
\bibliography{h}
\end{document}